\documentclass[aps,pra,reprint,superscriptaddress]{revtex4-2}

\usepackage{amsmath}
\usepackage{amssymb}
\usepackage{graphicx}
\graphicspath{ {fig/} }
\usepackage{mathtools}
\usepackage{xcolor}

\begin{document}


\title{Single-qubit quantum gate at an arbitrary speed}


\author{Seongjin Ahn}
\thanks{These authors contributed equally to this work.}
\affiliation{Department of Physics, KAIST, Daejeon 34141, Republic of Korea}

\author{Kichan Park}
\thanks{These authors contributed equally to this work.}
\affiliation{Department of Mathematical Sciences, KAIST, Daejeon 34141, Republic of Korea}

\author{Daehee Cho}
\affiliation{Department of Mathematical Sciences, KAIST, Daejeon 34141, Republic of Korea}

\author{Mikyoung Lim}
\email[]{hanmklim@kaist.ac.kr}
\affiliation{Department of Mathematical Sciences, KAIST, Daejeon 34141, Republic of Korea}

\author{Taeyoung Choi}
\email[]{tchoi@ewha.ac.kr}
\affiliation{Department of Physics, Ewha Womans University, Seoul 03760, Republic of Korea}

\author{Andrey S. Moskalenko}
\email[]{moskalenko@kaist.ac.kr}
\affiliation{Department of Physics, KAIST, Daejeon 34141, Republic of Korea}


\date{\today}

\begin{abstract}
	Quantum information processing comprises physical processes, which obey the quantum speed limit (QSL): high speed requires strong driving.
	Single-qubit gates using Rabi oscillation, which is based on the rotating wave approximation (RWA), satisfy this bound in the form that the gate time $T$ is inversely proportional to the Rabi frequency $\Omega$, characterizing the driving strength. However, if the gate time is comparable or shorter than the qubit period $T_{0} \equiv 2\pi / \omega_{0}$, the RWA actually breaks down since the Rabi frequency has to be large compared to the qubit frequency $\omega_{0}$ due to the QSL, which is given as $T \gtrsim \pi/\Omega$.
	We show that it is possible to construct a universal set of single-qubit gates at this strong-coupling and ultrafast regime, by adjusting the central frequency $\omega$ and the Rabi frequency $\Omega$ of the driving pulse.
	We observe a transition in the scaling behavior of the central frequency from the long-gate time regime ($T \gg T_{0}$) to the short-gate time ($T \ll T_{0}$) regime.
	In the former, the central frequency is nearly resonant to the qubit, i.e., $\omega \simeq \omega_{0}$, 
	whereas in the latter, the central frequency is inversely proportional to the gate time, i.e., $\omega \sim \pi/T$. 
	We identify the transition gate time at which the scaling exponent $n$ of the optimal central frequency $\omega \sim T^{n}$ changes from $n=0$ to $n=-1$.
	In the frequency domain, we find that the Fourier component of the driving pulse at the qubit frequency is nearly constant of $T$ and converges to the half of the gate angle in both long- and short-gate time limits.
\end{abstract}


\maketitle

%

\section{Introduction}

Precise operation of elementary quantum logic gates is essential for a fault-tolerant quantum information processing. Quantum error correction, which is integral for realizing the fault tolerance, requires at least a certain level of accuracy threshold of its constituent operations \cite{knill1996thresholdaccuracyquantumcomputation}. More importantly, even though the physical error rate is below the threshold, further improvement of accuracy leads to a faster exponential decay of the corresponding logical error rate as we put more resources to the error correction \cite{acharya2023}.
A recent realization of a below-threshold surface code suggests that around 50 qubits can be used instead of more than 100 qubits to achieve the same level of error correction, by improving the physical error rate approximately by a factor of two \cite{acharya2024quantumerrorcorrectionsurface}. Improving gate fidelity is an important part of reducing the physical error rate of the system.

For a precise quantum gate operation in the presence of environmental decoherence, the gate time $T$ should be sufficiently short compared to the coherence time of the system. For a given driving strength characterized by a Rabi frequency $\Omega$, the gate time is lower-bounded by the quantum speed limit (QSL) \cite{mandelstam1945, Aifer_2022},
\begin{equation}\label{eq:speed-limit}
	T \gtrsim \frac{\pi}{\Omega}.
\end{equation}
It says that the stronger you drive,
the faster you can finish the operation.
For a resonantly driven qubit, Rabi oscillation \cite{PhysRev.51.652,PhysRev.57.522} follows this relation as long as the driving strength $\Omega$ is sufficiently small compared to the qubit frequency $\omega_{0}$, so that the rotating wave approximation (RWA) holds \cite{cohen1998atom}.
For example, if the envelope of the driving pulse is constant in time, 
the Rabi oscillation requires a certain amount of driving time given as $T = \pi / \Omega$.
This implies, due to the weak driving condition $\Omega \ll \omega_{0}$ imposed by the RWA, the gate time should be sufficiently long compared to the qubit period $T_{0} \equiv 2\pi / \omega_{0}$.
If we drive the qubit stronger such that $\Omega \lesssim \omega_{0}$, the RWA starts to break down and the fidelity of the quantum gate decreases.
Perturbative approaches have been developed to identify the error terms,
including the Bloch-Siegert shift \cite{PhysRev.57.522} and its generalization to an arbitrary order with respect to $\Omega / \omega_{0}$ \cite{ZEUCH2020168327}.
The perturbative expansion can converge for moderate driving strength $\Omega \lesssim \omega_{0}$, and this can be used to push the gate time up to $T \gtrsim \pi / \Omega \gtrsim \pi / \omega_{0} \sim T_{0}$.
However, if the operation time were to be comparable or even shorter than the qubit period, i.e., $T \lesssim T_{0}$, then strong driving $\Omega \gtrsim \omega_{0}$ is required due to the speed limit, Eq.\,(\ref{eq:speed-limit}).
In this strong-coupling regime, the RWA and even its perturbative corrections are expected to break down.
This regime is particularly important for qubits with low frequencies such as tens of MHz, where a single-qubit period is on the order of $10\,\mathrm{ns}$ \cite{PhysRevX.9.041041,PhysRevX.10.041051,PhysRevX.11.011010,PRXQuantum.5.020326}, being slow compared to few-GHz qubits such as transmon or atomic hyperfine qubits by orders of magnitude.

To implement single-qubit quantum gates at this nonperturbative, ultrafast regime, a couple of exactly solvable two-level models, which are valid for an arbitrary driving strength, have been used. In particular, a spin-echo-type driving, which consists of two pulses with a delay in between, has been utilized to implement single-qubit gates on a trapped-ion hyperfine qubit \cite{PhysRevLett.105.090502} based on the Rosen-Zener model \cite{PhysRev.40.502} and a heavy-fluxonium qubit \cite{PhysRevX.11.011010} based on the finite-time Landau-Zener model \cite{PhysRevA.53.4288}. Both works have implemented two pulses with opposite polarity and individual pulse durations shorter than the qubit period $T_{0}$. Selecting $t=0$ at the center of the sequence, the temporal pulse shape $f(t)$, is anti-symmetric, i.e., $f(-t) = -f(t)$.
However, for such a spin-echo-based scheme the gate time has a lower bound which is constant with respect to $\Omega$, instead of being inversely proportional to $\Omega$ as Eq.\,(\ref{eq:speed-limit}).
For example, the gate time $T$ for the demonstrated $\pi$ gate in Ref.\,\cite{PhysRevLett.105.090502} is around $T_{0}/2 = \pi/\omega_{0}$. Comparing with Eq.\,(\ref{eq:speed-limit}), the gate time may be shortened if we drive stronger such that $\Omega \gtrsim \omega_{0}$. 
However, although each pulse gets stronger and thus shorter, the time delay between the two pulses is still necessary. For implementing the $\pi$ gate, the required delay is about $T_{0}/2$, being the lower bound of the gate time, no matter how large the driving strength $\Omega$ is.
This is because they drive in one direction, say, along the $x$ direction, and implement a rotation gate around another axis, say the $y$ direction, resulting in the QSL $T \ge T_{0}/2$ \cite{PhysRevLett.111.260501}. The equality holds when the two opposite-sign pulses are ly short and strong, i.e., the Dirac-delta-like pulses.
The above mentioned experiments realized nearly optimal driving in that the gate time $T$ almost saturated the QSL, namely, $T_{0}/2=\pi/\omega_{0}$ for the $R_{y}(\pi)$ gate \cite{PhysRevLett.105.090502}, and $T_{0}/4=\pi/2\omega_{0}$ for the $R_{y}(\pi/2)$ gate \cite{PhysRevX.11.011010}, determined by the qubit frequency $\omega_{0}$ and not by the driving strength $\Omega$ as in Eq.\,(\ref{eq:speed-limit}).

The QSL can be improved if we switch our target gate to the rotation around the same axis as the driving axis, i.e., the $x$ direction mentioned above. 
Then, a single pulse is enough to implement the gate within an arbitrary time $T > 0$, where the equality holds for a Dirac-delta pulse with a suitable strength to match the rotation angle of the gate \cite{PhysRevLett.111.260501}. 
However, such a pulse is an idealization since, in practice, the driving strength and/or the time resolution are limited.
In addition, a too short pulse may lead to 
information leakage outside of the qubit subspace \cite{PhysRevLett.103.110501,PhysRevA.94.032306}. 
Instead, it is possible to approximate the delta function with an ultrashort ($T \ll T_{0}$) and strong ($\Omega \gg \omega_{0}$) pulse, which has been realized with a nonzero error \cite{PhysRevLett.105.090502,PhysRevX.11.011010}. For example, in Ref.\,\cite{PhysRevLett.105.090502}, a pulse with $T = 14.8\,\mathrm{ps} < 79.1\,\mathrm{ps} = T_{0}$ $(\omega_{0} / 2\pi = 12.6\,\mathrm{GHz})$ has been implemented with a temporal shape of approximately that of a hyperbolic secant, $f(t) \propto \mathrm{sech}(\pi t/T)$, yielding 72\% fidelity for a $\pi$ gate.
The Rosen-Zener model predicts that the fidelity becomes unity if and only if $T \rightarrow 0$ for the given pulse shape, which is not feasible in practice. 

To find a good pulse shape with a short ($T \ll T_{0}$), but not infinitesimally short, duration $T$, several optimal control frameworks have been applied. The underlying idea is that the pulse shape is parameterized and the gate fidelity is optimized with respect to the parameters. For example, chopped random basis (CRAB) algorithm \cite{PhysRevLett.106.190501} with 15 parameters which corresponds to a five-frequency subset of the Fourier basis has been used to find optimal pulse shape beyond RWA. The technique obtained 99.86\% theoretical fidelity and 99\% experimental fidelity for a $\pi$ gate on a single nitrogen-vacancy center \cite{Scheuer_2014}. Using more parameters can improve fidelity. 
However, experimental realization benefits from having fewer parameters, making calibration more feasible.
A two-parameter driving pulse has been suggested with theoretical fidelity more than 99.9\% for $\Omega \sim \omega_{0}$ \cite{Yudilevich_2023}.

%

Further, optimal pulse shape which gives unit fidelity has been found within an analytical approach based on the optimal control theory. 
There, considering the Schr\"{o}dinger equation for the propagator $U(t,t_{i})$ with a fixed initial operator $U(t_{i},t_{i}) = I$ and a fixed target gate $U(t_{f},t_{i}) = U_{\mathrm{target}}$, a functional such as energy of the driving pulse, namely, $\int {dt [\Omega f(t)]^{2}}$, is minimized among various trajectories connecting $I$ and $U_{\mathrm{target}}$.
For example, a $y$-axis rotation gate, $R_{y}(\theta_{g})$, can be implemented with a single $x$-axis driving, i.e., $H(t) = \hbar\omega_{0}\sigma_{z}/2 + \Omega f(t) \sigma_{x}$, achieving unit fidelity. Among various pulse shapes that generate the target gates, the pulse shape with minimal energy is found to be a Jacobi elliptic function \cite{D-Alessandro2001}. 
The function has few parameters and it makes the experimental calibration of the control pulse relatively easy.
The function, however, has discontinuities at the start and/or at the end of the pulse, so that the actual generated shape of pulse may be affected by a finite bandwidth of the pulse generator \cite{PhysRevLett.106.190501}. Thus, it might be preferable to constrain the pulse shape to be smooth in the time domain including at the start and the end of the pulse, such as ensured by a Gaussian pulse envelope. Then the pulse can have a relatively small bandwidth for the given pulse duration \cite{BAUER1984}. This helps reducing pulse distortions after band pass filters \cite{PhysRevA.96.022330} as well as leakage errors outside of the qubit subspace \cite{PhysRevLett.116.020501}.

As an alternative approach, 
unit fidelity can also be achieved based on an adiabatic passage and its shortcuts to it \cite{RevModPhys.91.045001}. 
For a given initial and a target state, adiabatic theorem assures population transfer between them as long as the state evolution is sufficiently slow. 
In practice, the speed should be away from zero, i.e., the speed cannot be infinitesimally low. The corresponding error occurs due to diabatic transitions.
By adding a so-called counterdiabatic (CD) driving, it is possible to prevent the diabatic transitions and achieve population transfer with unit fidelity at an arbitrary speed \cite{Demirplak_Rice_2003,Berry_2009}.
For a qubit, the CD driving corresponds to additional component, $\sigma_{y}$, of the Hamiltonian.
In experiment, the CD driving can be realized, under the RWA, by implementing time-dependent central frequency $\omega(t)$, i.e., chirp, and phase $\phi(t)$ of the pulse \cite{PhysRevLett.110.240501}.

We show that there is an alternative way to implement single-qubit quantum gates at arbitrary speed by tuning the central frequency $\omega$, phase $\phi$, and maximum amplitude $\Omega$ of the pulse, where all three parameters are constant over time.
The paper is organized as follows. In Sec.\,\ref{sec:model-single-qubit-gates}, we introduce the model Hamiltonian and sketch the RWA-based approach to analytically find an optimal pulse shape in the long-$T$ ($T \gg T_{0}$) regime. Then, we go to the opposite regime where the gate time is much shorter than the qubit period, i.e., $T \ll T_{0}$.
We discuss the limit of vanishing gate time ($T\rightarrow 0$) and how the error grows as the gate time increases. 
We identify the leading-order error in terms of the gate time so that we can search for an optimal pulse shape to correct the dominant part of the error.
We show that in this regime the leading-order error (i.e., the gate infidelity) is nonzero for a rotational angle $\theta_{g} \in [0,2\pi]$, if the pulse shape is nonnegative, i.e., $f(t) \ge 0$.
In order to remove the error, we introduce sufficient negativity of the pulse by increasing (blue-detuning) the central frequency from the qubit frequency, i.e., $\omega > \omega_{0}$.
By using the set of analytically obtained parameters as an initial guess, we perform numerical optimization and show that a set of parameters achieving unit fidelity up to the numerical precision exist for all regimes of the gate time.
The optimal central frequency $\omega$ approaches the qubit frequency in the long-$T$ limit but scales inversely proportional to the gate time in the short-$T$ limit. 
In the intermediate regime, there occurs a transition in the scaling behavior of $\omega$.
By finding where the two asymptotic solutions intersect, 
the gate time at which such transition occurs is identified.
In Sec.\,\ref{sec:spectra}, we analyze frequency spectra of optimal pulse shapes for different gate times.
We show that the Fourier component of the optimal pulse shape at the qubit frequency $\omega_{0}$ converges to the same value, namely, the half of the gate rotation angle $\theta_{g}/2$, in both limits. It is conserved up to few percents in all considered regimes.
In Sec.\,\ref{sec:rot-frame-magnus}, 
in order to understand the few-percent deviation of the resonant Fourier component from $\theta_{g}/2$ in the intermediate regime, $T \sim T_{0}$, 
we expand the exponent of the unitary operator with respect to $\Omega T \sim \theta_{g}$.
We show that the deviation comes predominately from the third-order term in the expansion.
In Sec.\,\ref{sec:universal-set}, we explain how to construct a universal set of single-qubit gates based on the optimal parameters obtained in Sec.\,\ref{sec:model-single-qubit-gates}.
In Sec.\,\ref{sec:conclusion}, we conclude this work with remarks on potential experimental noises which may affect the gate fidelity.

\section{\label{sec:model-single-qubit-gates}Multicycle and subcycle single-qubit gates}

A qubit driven by an external field can be modeled as $H_{\mathrm{lab}}(t) = H_{0} + V(t)$ where $H_{0} = \hbar \omega_{0} \sigma_{z}/2$ describes the free evolution of the qubit and $V(t) = \hbar \Omega f(t) \sigma_{x}$ induces transitions between the two levels. $\Omega$ is the maximum amplitude of the coupling between the qubit and the external driving field such that the pulse shape $f(t)$ is normalized as $\max_{t}|f(t)|=1$. For the two energy eigenstates of the qubit, denoted as $|0\rangle$ and $|1\rangle$, $\sigma_{z}=|1\rangle\langle 1| - |0\rangle\langle 0|$.
To see the effect of the pulse, we introduce the rotating frame such that
\begin{equation}\label{eq:Ulab-U0-Urot-U0}
	U_{\mathrm{lab}}(t,-T/2) \equiv U_{0}(t,0)U_{\mathrm{rot}}(t,-T/2;0)U_{0}(0,-T/2),
\end{equation}
where $t \in [-T/2,T/2]$ for the total gate time $T$. The lab-frame propagator satisfies $\dot{U}_{\mathrm{lab}}=-(i/\hbar)H_{\mathrm{lab}}U_{\mathrm{lab}}$ and the free propagator $U_{0}(t,t') = \exp[-(i/\hbar)(t-t')H_{0}] = \exp[-i\omega_{0} (t-t')\sigma_{z}/2] = R_{z}[\omega_{0} (t-t')]$ is an operator for rotation about the $z$-axis.
The Hamiltonian in the rotating frame can be written as
\begin{equation}\label{eq:H-rot}
	\begin{split}
		H_{\mathrm{rot}}(t) 
		&= U_{0}^{\dagger}(t,0) \left[H_{\mathrm{lab}}(t) - H_{0}\right] U_{0}(t,0)
		\\
		&= \hbar \Omega f(t) [\cos{(\omega_{0} t)} \sigma_{x} - \sin(\omega_{0} t) \sigma_{y}]
	\end{split}
\end{equation}
from $\dot{U}_{\mathrm{rot}} = -(i/\hbar)H_{\mathrm{rot}}U_{\mathrm{rot}}$ and Eq.\,(\ref{eq:Ulab-U0-Urot-U0}).
Let us consider a pulse shape 
\begin{equation}\label{eq:pulse-shape-ft}
	f(t) = f_{0}(t)\cos(\omega t + \phi),
\end{equation}
where $f_{0}(t)$ is an envelope function with an (effective) duration $\tau_{d}$. 
It is useful to introduce a normalized time $u \equiv t / \tau_{d} \in [-T_{d}/2,T_{d}/2]$ with $T_{d}\equiv T / \tau_{d}$. For example, a Gaussian pulse envelope can be then written as $f_{0}(t) = \exp[-(2t/\tau_{d})^{2}] = \exp[-(2u)^{2}] = f_{0}(u)$.

\begin{figure}
	\centering
	\includegraphics[width=1\linewidth]{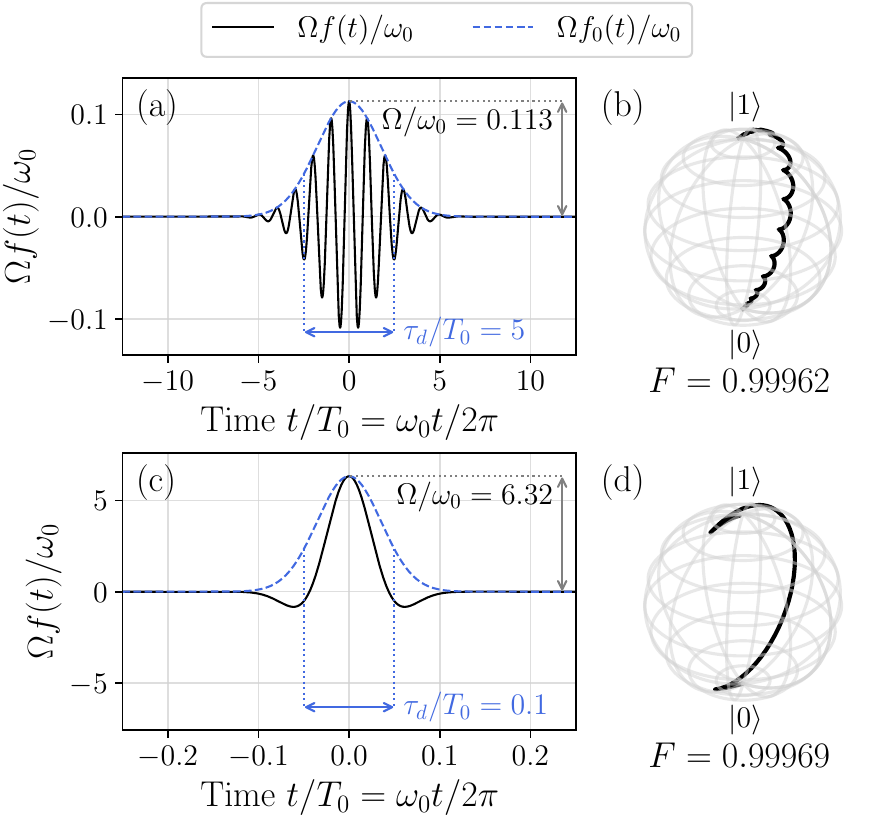}
	\caption{
		Driving pulse shapes for $R_{x}(\pi)$ gate and the corresponding trajectories of the Bloch vectors driven by (a,b) a multicycle pulse ($\tau_{d} = 5\,T_{0}$) and (c,d) a subcycle pulse ($\tau_{d}=0.1\,T_{0}$). For the multicycle pulse, the driving is resonant ($\omega=\omega_{0}$) and its driving strength $\Omega = 0.113\,\omega_{0}$ is determined by Eq.\,(\ref{eq:Omega-tau-d-rwa}).
		For the subcycle pulse, the central frequency $\omega=\varphi/\tau_{d}=5.72\,\omega_{0}$ and the driving strength $\Omega = 6.32\,\omega_{0}$ are determined by Eqs.\,(\ref{eq:varphi-omegataud}) and (\ref{eq:Omega-tau-d-subcycle}), respectively.
		In (a) and (c), the black solid lines represent the pulse shape $f(t)$ given as Eq.\,(\ref{eq:pulse-shape-ft}) with $\phi=0$, and the blue dashed lines indicate the envelope function $f_{0}(t) = \exp[-(2t/\tau_{d})^{2}]$. For both pulses, the ratio between the total gate time and the effective pulse duration is fixed to be $T/\tau_{d} = 5$. In (b) and (d), 
		$F$ indicates the corresponding gate fidelity.
		Since the pulse parameters are analytically obtained based on perturbation theories, either in terms of $\Omega/\omega_{0}\ll 1$ or $\tau_{d}/T_{0}\ll 1$, there exists nonzero infidelity $1-F$ due to the omitted higher-order terms. Parameters achieving unit fidelity up to numerical precision are obtained through numerical optimization of fidelity and are shown in Fig.\,\ref{fig:average-gate-fidel-and-parameters}.
	}
	\label{fig:pulse-shapes-and-bloch-vectors}
\end{figure}

A resonant pulse, i.e., a pulse with $\omega = \omega_{0}$, approximately generates a rotation about the axis defined as $\mathbf{n}_{\phi} \equiv \cos{\phi}\,\mathbf{e}_{x} + \sin{\phi}\,\mathbf{e}_{y}$, which is in the $xy$ plane. The rotational angle is $\theta_{0}(t,-T/2) \equiv \theta_{0}(t) \equiv \Omega \int_{-T/2}^{t}{dt' f_{0}(t')}$. 
This is valid under the rotating wave approximation (RWA) if the driving strength is weak such that $\Omega \ll \omega_{0}$. 
$R_{x}(\theta)$ and $R_{y}(\theta)$ can be realized by selecting $\phi=0$ and $\pi/2$, respectively. 
In Figs.\,\ref{fig:pulse-shapes-and-bloch-vectors}(a) and \ref{fig:pulse-shapes-and-bloch-vectors}(b), a driving pulse with strength $\Omega = 0.113\,\omega_{0}$ and duration $\tau_{d} = 5\,T_{0}$ is shown with the corresponding Bloch vector trajectory $\langle \pmb{\sigma} \rangle(t)$ driven by the pulse from the ground state, where $\pmb{\sigma} \equiv (\sigma_{x},\sigma_{y},\sigma_{z})$.
Given a target gate rotational angle, denoted as $\theta_{g}$, the condition of weak driving requires a long pulse duration. Precisely, $\tau_{d}$ should be much longer than the single period of the free rotation of Bloch sphere, $T_{0}$.
This can be seen from the requirement $\theta_{0}(T/2) = \theta_{g}$, which yields
\begin{equation}\label{eq:Omega-tau-d-rwa}
	\Omega\tau_{d} = \theta_{g} / s_{0},
\end{equation}
where $s_{0} = \frac{1}{\tau_{d}}\int_{-T/2}^{T/2}{dt f_{0}(t)} = \int_{-T_{d}/2}^{T_{d}/2}{du f_{0}(u)}$. 
Since the effective pulse area $\Omega \tau_{d}$ is fixed, a weak driving implies $\tau_{d} \sim \pi/\Omega \gg \pi/\omega_{0} \sim T_{0}$, i.e., the pulse should be sufficiently long compared to the duration of a single free rotation of the qubit.
However, the weak driving condition should be relieved to implement a faster gate due to the quantum speed limit \cite{mandelstam1945, Aifer_2022}. Then, the RWA breaks down and its correction for $\Omega \lesssim \omega_{0}$ has been developed \cite{PhysRev.57.522,ZEUCH2020168327}.

In the limit of a very short and strong driving pulse, such that $\tau_{d} \ll T_{0}$ while keeping the effective pulse area $\Omega \tau_{d}$ constant, the pulse $f(t)$ is nearly zero except for $|\omega_{0} t| \lesssim \omega_{0} \tau_{d} \ll 2\pi$ so that $\cos(\omega_{0} t) \simeq 1$ and $\sin(\omega_{0} t) \simeq 0$ to the leading-order approximation.
Hence, the rotating-frame Hamiltonian, Eq.\,(\ref{eq:H-rot}), can be approximated as
\begin{align*}
	H_{\mathrm{rot}}(t) \xrightarrow{\tau_{d} \ll T_{0}}
	\hbar\Omega f(t) \sigma_{x}
	= V(t),
\end{align*}
for $|t|\lesssim \tau_{d}$. 
This can be expected since the pulse shape $f(t)$ is modulated at a much faster timescale than $T_{0}$. Then the qubit dynamics is dominated by the driving term $V(t)$, which generates the corresponding propagator
\begin{equation*}
	\begin{split}
		U_{V}(t,-T/2) 
		&= \exp\left[-\frac{i}{\hbar}\int_{-T/2}^{t}{dt' V(t')}\right] 
		\\
		&= R_{x}[\theta(t,-T/2)],
	\end{split}
\end{equation*}
where $\theta(t,-T/2) \equiv \theta(t) \equiv 2\Omega \int_{-T/2}^{t}{dt' f(t')}$.
This approximation becomes more accurate as the pulse duration gets shorter, and in the limit of $\tau_{d} \rightarrow 0$, the rotating-frame propagator $U_{\mathrm{rot}}$ given as Eq.\,(\ref{eq:Ulab-U0-Urot-U0}) is expected to uniformly converge towards $U_{V}$ on the normalized time domain $u \in [-T_{d}/2,T_{d}/2]$. 
Although this can be a good approximation in the limit of ultrafast driving ($\tau_{d} \ll T_{0}$), it is not exact as long as the pulse duration is infinitesimally short. 
In order to estimate the deviation of $U_{V}$ from $U_{\mathrm{rot}}$ for a short but nonzero pulse duration, 
we move to the second interaction picture, following a decomposition suggested in Ref.\,\cite{MOSKALENKO20171}. The interaction picture is defined by
\begin{equation}\label{eq:short-and-strong-picture}
	\begin{split}
		&U_{\mathrm{rot}}(t,-T/2;0) 
		\\
		&\equiv U_{V}(t,0) \, \mathcal{U}(t,-T/2;0) U_{V}(0,-T/2)
		\\
		&= R_{x}[\theta(t,0)] \, \mathcal{U}(t,-T/2;0) R_{x}[\theta(0,-T/2)].
	\end{split}
\end{equation}
We expand the exponent of the propagator $\mathcal{U}$ in terms of a relative (effective) pulse duration defined as $\alpha \equiv \tau_{d}/T_{0} = \omega_{0} \tau_{d}/2\pi$.
The generating Hamiltonian $\mathcal{H}$ for $\mathcal{U}$, which satisfies $\dot{\mathcal{U}} = (-i/\hbar)\mathcal{H}\mathcal{U}$, can be written as
\begin{equation*}
	\begin{split}
		\mathcal{H}(t) 
		&= U_{V}^{\dagger}(t,0) [H_{\mathrm{rot}}(t) - V(t)] U_{V}(t,0)\\
		&\equiv \hbar \Omega f(t)\,\mathbf{a}(t)\cdot \pmb{\sigma}/2,
	\end{split}
\end{equation*}
where $\mathbf{a} \equiv (a_{x},a_{y},a_{z})$ for $a_{x}(t) = 2 \{\cos(\omega_{0} t) - 1\}$, 
$a_{y}(t) = -2\sin(\omega_{0} t)\cos{\theta(t,0)}$, and $a_{z}(t) = 2\sin(\omega_{0} t)\sin{\theta(t,0)}$.
Once again, since $f(t)$ is nearly zero except for a short duration $\tau_{d}$ around $t=0$, such that $|\omega_{0} t| \lesssim |\omega_{0} \tau_{d}| \ll 2\pi$, we expect $a_{x}(t) = \mathcal{O}(\alpha^{2})$ and $a_{y}(t),a_{z}(t) = \mathcal{O}(\alpha^{1})$. 
This implies $\mathcal{U} \rightarrow I$ as $\alpha \rightarrow 0$, where $I$ is the identity operator of SU(2).
To estimate the magnitude of the first-order error, we perform the Magnus expansion \cite{Magnus,BLANES2009151} and sort out $\mathcal{O}(\alpha^{1})$ terms, which gives
\begin{equation*}
	\mathcal{U}(t,-T/2;0) = \exp\left[-i \alpha A^{(1)}(t) + \mathcal{O}(\alpha^{2})\right],
\end{equation*}
where $A^{(1)}(t) = A^{(1)}_{1y}(t)+A^{(1)}_{1z}(t)$ consists of $A^{(1)}_{1y}(t) = -2\pi \Omega \tau_{d} s_{1c}(t) \sigma_{y}$ and $A^{(1)}_{1z}(t) = 2\pi \Omega \tau_{d} s_{1s}(t) \sigma_{z}$ 
for $s_{1c}(t) + i s_{1s}(t)\equiv \int_{-T_{d}/2}^{t/\tau_{d}}{du f(u) u \exp{[i\theta(u,0)]}}$.
Targeting $U_{\mathrm{rot}}(T/2,-T/2;0) = R_{x}(\theta_{g})$ and using Eq.\,(\ref{eq:short-and-strong-picture}) with $\theta(T/2) \equiv \theta(T/2,-T/2) = \theta(T/2,0) + \theta(0,-T/2)$, we get
\begin{equation*}
	[\theta_{g} - \theta(T/2)]\frac{\sigma_{x}}{2} = \alpha A^{(1)}(T/2) + \mathcal{O}(\alpha^{2}),
\end{equation*}
which is satisfied up to $\mathcal{O}(\alpha^{1})$ by fulfilling the following conditions:
\begin{subequations}\label{eq:cond-subcycle-regime}
	\begin{align}
		\label{eq:theta-theta_x}
		\theta(T/2) &= \theta_{g},\\
		\label{eq:s1c-zero}
		s_{1c}(T/2) &= 0,\\
		\label{eq:s1s-zero}
		s_{1s}(T/2) &= 0.
	\end{align}
\end{subequations}
The first condition, Eq.\,(\ref{eq:theta-theta_x}), requires the integral of the pulse, i.e., $\int_{-T/2}^{T/2}{dt \,\Omega f(t)}$, to be $\theta_{g}/2$, which is half of the target rotational angle. 
This condition can be satisfied by scaling the effective pulse area to be
\begin{equation}\label{eq:Omega-tau-d-subcycle}
	\Omega \tau_{d} = \theta_{g} / 2s,
\end{equation}
provided that the integral of the shape function $s \equiv \int_{-T_{d}/2}^{T_{d}/2}{du f(u)}$ is nonzero.
Comparing this with Eq.\,(\ref{eq:Omega-tau-d-rwa}), we note that in the multicycle limit ($\alpha \gg 1$), where the RWA is valid due to the weak and resonant driving, the integral of the \textit{envelope} of the pulse, i.e., $\int_{-T/2}^{T/2}{dt \,\Omega f_{0}(t)}$ should be $\theta_{g}$, without the factor of $1/2$ due to the absence of the counter-rotating term dropped in the RWA. 
To satisfy Eqs.\,(\ref{eq:s1c-zero}) and (\ref{eq:s1s-zero}), we note that one of the integrals vanishes 
if $f(u)$ is even or odd,
because, in this case, $\theta(u,0) = 2\Omega \tau_{d} \int_{T_{d}/2}^{u}{du' f(u')}$ has the parity opposite to that of $f(u)$. 
If $f(u)$ is odd, the integral of $f(u)$ vanishes, ending up with $\theta(T/2) = 0 \neq \theta_{g}$.
Thus, $f(u)$ should be even in order to satisfy Eq.\,(\ref{eq:theta-theta_x}) for a nonzero $\theta_{g}$.
In this case, $\theta(u,0)$ is odd, and thus, $s_{1c}(T/2) = 0$, automatically satisfying Eq.\,(\ref{eq:s1c-zero}).
Now we are left with Eq.\,(\ref{eq:s1s-zero}), which can be written as
\begin{equation}\label{eq:s1s-half-even}
	\int_{0}^{T_{d}/2}{du f(u) u \sin\theta(u,0)} = 0,
\end{equation}
where we have used that the integrand is even.
The angle $\theta(u,0)$ depends on $\Omega \tau_{d}$ but it can be written purely in terms of $f(u)$, by using Eq.\,(\ref{eq:theta-theta_x}), as
\begin{equation}\label{eq:theta-u-0}
	\theta(u,0) = \frac{\theta_{g}}{2}\frac{s(u,0)}{s(T_{d}/2,0)},
\end{equation}
where $s(u,0) = \int_{0}^{u}{du' f(u')}$.
There are more than one possible $f(u)$ that satisfy Eq.\,(\ref{eq:s1s-half-even}).
However, such $f(u)$ should change its sign at least once for $u > 0$. Otherwise, $f(u)$ is nonnegative for all $u$, $\theta(u,0)$ is monotonic and has its upper bound $\theta(T_{d}/2,0) = \theta(T_{d}/2,-T_{d}/2) / 2 = \theta_{g} / 2$. Thus, for a gate with its rotational angle up to $2\pi$, we have $\theta(u,0) \le \theta(T_{d}/2,0) \le \pi$, for which $\sin\theta(u,0)$ is nonnegative and the integral in Eq.\,(\ref{eq:s1s-half-even}) is strictly positive, so that Eq.\,(\ref{eq:theta-u-0}) cannot be satisfied.


There are several ways to introduce a negativity in $f(u)$. We show that a nonnegative envelope $f_{0}(u)$ multiplied by a cosine function, Eq.\,(\ref{eq:pulse-shape-ft}), which was used in the multicycle regime, can also be adapted in the subcycle regime, by choosing suitable parameters, namely, $\Omega$, $\omega$, and $\phi$. 
First, for a given pulse duration $\tau_{d}$, the driving strength $\Omega$ can be determined according to Eq.\,(\ref{eq:Omega-tau-d-subcycle}), satisfying Eq.\,(\ref{eq:theta-theta_x}).
Second, we set $\phi = 0$ for making the pulse shape even, in order to satisfy Eq.\,(\ref{eq:s1c-zero}).
Last, the central frequency $\omega$ can be used to fulfill Eq.\,(\ref{eq:s1s-zero}), which translates into Eq.\,(\ref{eq:s1s-half-even}) for an even pulse.
We note that $f(u) = f_{0}(u) \cos(\omega \tau_{d} u)$ and $\theta(u,0)$ are parameterized by a single quantity, namely, $\omega \tau_{d}$, as can be seen from Eq.\,(\ref{eq:theta-u-0}). Among $\omega \tau_{d}$ which satisfies Eq.\,(\ref{eq:s1s-half-even}), we take the smallest possible value
\begin{equation}\label{eq:varphi-omegataud}
	\varphi \equiv \min\{\omega \tau_{d} : s_{1s}(T/2) = 0\}.
\end{equation}
For a given pulse duration $\tau_{d}$, minimizing $\omega$ keeps the integral $s$ far from zero, allowing the effective pulse area $\Omega \tau_{d}$, determined by Eq.\,(\ref{eq:Omega-tau-d-subcycle}), to be finite.
For a given target gate angle $\theta_{g}$, the minimum $\omega \tau_{d}$, i.e., $\varphi = \varphi[f_{0}]$, is a characteristic quantity of the pulse envelope function $f_{0}(u)$. For example, the Gaussian envelope has $\varphi/2\pi = 0.572$ for $\theta_{g} = \pi$.
In Figs.\,(\ref{fig:pulse-shapes-and-bloch-vectors}c) and (\ref{fig:pulse-shapes-and-bloch-vectors}d), a pulse with a subcycle duration ($\tau_{d} = 0.1\,T_{0}$) is shown with the corresponding Bloch vector trajectory. 
The strength and the central frequency of the pulse are given as $\Omega = 6.32\,\omega_{0}$ and $\omega = \varphi/\tau_{d} = 5.72\,\omega_{0}$, respectively, in order to satisfy Eq.\,(\ref{eq:cond-subcycle-regime}).


\begin{figure}
	\centering
	\includegraphics[width=\linewidth]{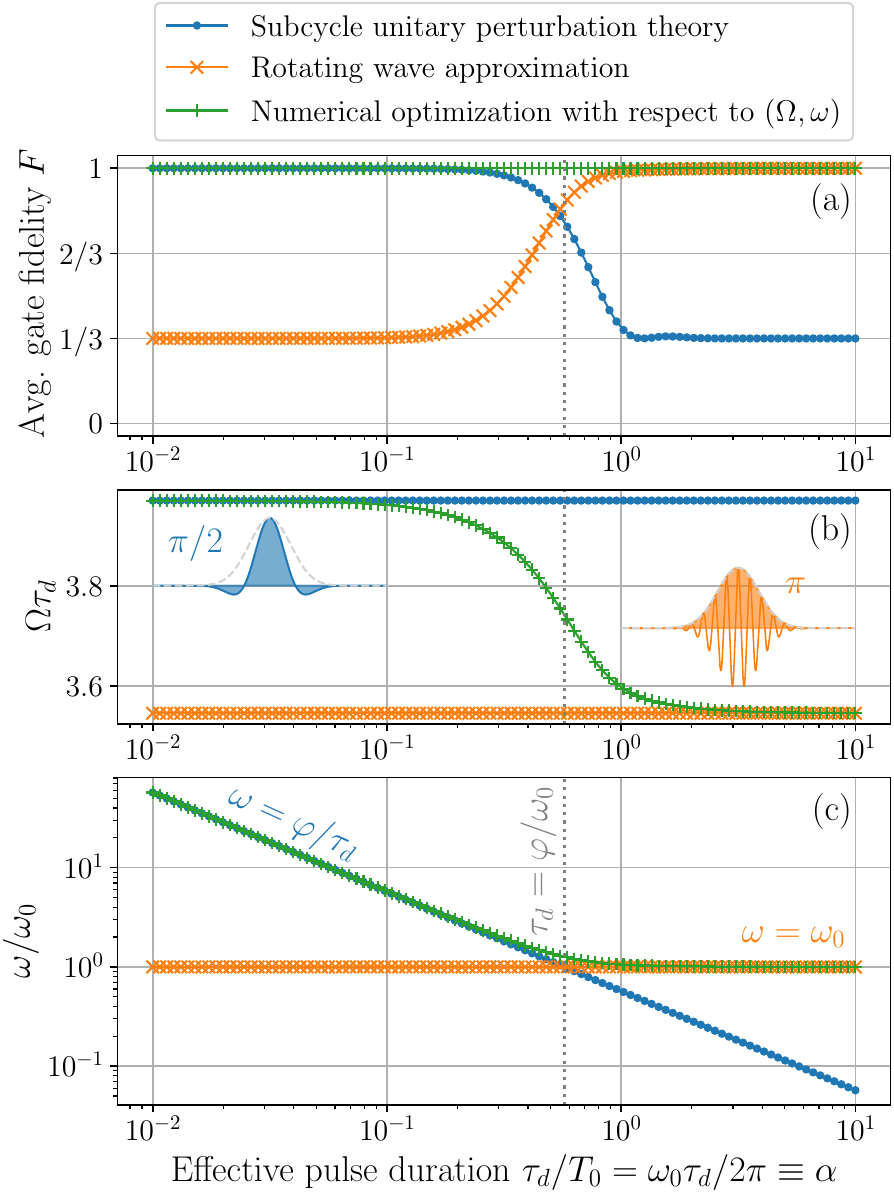}
	\caption{
		Average gate fidelity and the corresponding pulse parameters. The pulse shape is given by Eq.\,(\ref{eq:pulse-shape-ft}) with $\phi = 0$. The blue line with circles, orange line with `x', and the green line with `+' markers denote the results from the subcycle unitary perturbation theory [Eqs.\,(\ref{eq:Omega-tau-d-subcycle}) and (\ref{eq:varphi-omegataud})], rotating wave approximation [Eq.\,(\ref{eq:Omega-tau-d-rwa}) and $\omega = \omega_{0}$], and the numerical optimization of fidelity as a function of $\Omega$ and $\omega$, respectively. (a) Average gate fidelity $F$ given as Eq.\,(\ref{eq:average-gate-fidelity}). For each pulse duration, the total gate time is given as $T = 5\tau_{d}$. The numerical optimization gives unit fidelity ($F=1$) up to numerical accuracy. (b) Effective pulse area $\Omega\tau_{d}$. (c) Central frequency $\omega$ normalized by the qubit frequency $\omega_{0}$. Gray dotted vertical lines show the pulse duration at the transition from the subcycle to the multicycle regime, where the exponent of $\tau_{d}$ in the expression for $\omega \sim \tau_{d}^{n}$ changes from $n=-1$ to $n=0$. 
	}
	\label{fig:average-gate-fidel-and-parameters}
\end{figure}

In Fig.\,\ref{fig:average-gate-fidel-and-parameters}, we compare the pulse parameters for $R_{x}(\pi)$ computed from the subcycle unitary perturbation theory, i.e., Eqs.\,(\ref{eq:Omega-tau-d-subcycle}) and (\ref{eq:varphi-omegataud}), to those from the rotating wave approximation, namely, Eq.\,(\ref{eq:Omega-tau-d-rwa}) and $\omega = \omega_{0}$. 
In Fig.\,\ref{fig:average-gate-fidel-and-parameters}(a), we show the average gate fidelity,
which is a state fidelity between the target state $R_{x}(\theta_{g})|\psi(-T/2)\rangle$ and the final state $|\psi(T/2)\rangle = U_{\mathrm{rot}}(T/2,-T/2;0)|\psi(-T/2)\rangle$, averaged over all possible initial states $|\psi(-T/2)\rangle$.
The average gate fidelity has an analytical expression \cite{PhysRevA.60.1888,NIELSEN2002249}
\begin{equation}\label{eq:average-gate-fidelity}
	F = \frac{1}{3} + \frac{2}{3} F_{e},
\end{equation}
where the entanglement fidelity $F_{e} = |\mathrm{Tr}[R_{x}^{\dagger}(\theta_{g}) U]/2|^{2}$ is $1$ if and only if the generated propagator $U = U_{\mathrm{rot}}(T/2,-T/2;0) \in \mathrm{SU(2)}$ is identical to the target gate $R_{x}(\theta_{g})$.
As we see, the RWA works well in the multicycle regime ($\tau_{d} \gg T_{0}$) and breaks down as the pulse duration gets shorter. The subcycle unitary perturbation theory works in the opposite limit ($\tau_{d} \ll T_{0}$), but it breaks down in the multicycle regime. 
Since the same ansatz, Eq.\,(\ref{eq:pulse-shape-ft}), is used for the pulse shape in both the subcycle and the multicycle regime, with appropriate parameters, $\Omega$ and $\omega$, we may find their optimal values for the intermediate regime where the pulse duration $\tau_{d}$ is comparable to the qubit timescale $T_{0}$. To search for the optimal values of the parameters, we minimize the average gate infidelity $1 - F$ with respect to these parameters. We are able to find the optimal values which give numerically exact fidelity $F = 1$ as shown in Fig.\,\ref{fig:average-gate-fidel-and-parameters}(a) by the green solid line with vertical markers. In Figs.\,\ref{fig:average-gate-fidel-and-parameters}(b) and \ref{fig:average-gate-fidel-and-parameters}(c) we show the optimal values of $\Omega$ and $\omega$ for a range of pulse durations $10^{-2}\,T_{0} \le \tau_{d} \le 10^{1}\,T_{0}$. The optimal values of $\Omega$ and $\omega$ converge to the parameters of the subcycle (multicycle) regime as the pulse duration gets shorter (longer) than the qubit timescale $T_{0}$.
We note that the effective pulse area, $\Omega \tau_{d}$, is constant in both regimes, which implies $\Omega$ is inversely proportional to $\tau_{d}$ in those regimes. However, the central frequency $\omega$, which determines the pulse shape through Eq.\,(\ref{eq:pulse-shape-ft}), has a different behavior from that of $\Omega$. In the subcycle regime, $\omega$ is inversely proportional to $\tau_{d}$, just like $\Omega$, so that $\omega \tau_{d}$ is fixed. Thus, the pulse shape does not change except the scaling in time, 
as can be seen from $f(t/\tau_{d}) = f(u) = e^{-(2u)^{2}}\cos(\omega \tau_{d} u)$. However, in the multicycle regime, the central frequency converges to the qubit frequency $\omega_{0}$, with little dependence on the pulse duration. As shown in Fig.\,\ref{fig:average-gate-fidel-and-parameters}(c), the transition in the scaling behavior from $\omega = \varphi / \tau_{d} \sim \tau_{d}^{-1}$ (blue line with dots) to $\omega = \omega_{0} \sim \tau_{d}^{0}$ (orange line with `x' markers) occurs at the crossing point $\tau_{d} = \tau_{0}$, where
\begin{equation}\label{eq:tau0}
	\tau_{0} \equiv \varphi / \omega_{0} = \frac{\varphi}{2\pi}T_{0}
\end{equation} 
and $\varphi$ is given by Eq.\,(\ref{eq:varphi-omegataud}). 
In Table \ref{table:tau0}, we list numerical values of $\tau_{0}$ for different pulse envelope functions and gate angles. We note that $\tau_{0}$ depends on how the pulse duration $\tau_{d}$ is defined for 
a given pulse envelope function $f_{0}(t)$. For example, we can scale the pulse duration by $\tau_{d} \rightarrow 2\sqrt{2}\,\tau_{d}$ so that the Gaussian envelope becomes $e^{-(2t/\tau_{d})^{2}} \rightarrow e^{-(t/\sqrt{2}\tau_{d})^{2}}$, which has been used for standard quantum gates \cite{PhysRevLett.103.110501,PhysRevLett.116.020501}. Since the transition point $\tau_{0}$ also gets scaled, however, what determines the speed regime is 
$\tau_{d}/\tau_{0}$, which is invariant under the scaling of $\tau_{d}$.
We also note that $\alpha \equiv \tau_{d}/T_{0}$ is not invariant under the scaling of $\tau_{d}$.
However, since the transition pulse duration is on the same order as the single qubit period, i.e., $\tau_{0} \sim T_{0}$, $\alpha=1$ gives a good estimate of the transition point between the subcycle and multicycle regimes.
In Figs.\,\ref{fig:average-gate-fidel-and-parameters}(c), \ref{fig:spectra-and-envelope-comparison}(b), and \ref{fig:spectra-and-envelope-comparison}(c), we demonstrate that $\tau_{d}/\tau_{0} = 1$ serve as a more quantitative criterion to characterize the boundary between the two regimes.

 \begin{table}
	\caption{\label{table:tau0} Transition value $\tau_0$ defined as Eq.\,(\ref{eq:tau0}) for different envelope functions $f_{0}(t)$ and gate angles $\theta_{g}$.}
	\begin{ruledtabular}
		\begin{tabular}{c c c c}
			$f_{0}(t)$ & $T$ & $\theta_{g}$ & $\tau_{0}$ \\
			\hline
			$1$ & $\tau_{d}$ & $\pi$ & $0.7328\,T_{0}$ \\
			$1 - |t/\tau_{d}|$ & $2\tau_{d}$ & $\pi$ & $0.4899\,T_{0}$ \\
			$\mathrm{sech}(\pi t/\tau_{d})$ & $5\tau_{d}$ & $\pi$ & $0.4261\,T_{0}$ \\
			$e^{-(2t/\tau_{d})^{2}}$ & $5\tau_{d}$ & $\pi$ & $0.5718\,T_{0}$ \\
			$e^{-(2t/\tau_{d})^{2}}$ & $5\tau_{d}$ & $\pi/2$ & $0.5534\,T_{0}$ \\
			$e^{-(2t/\tau_{d})^{2}}$ & $5\tau_{d}$ & $\pi/4$ & $0.5498\,T_{0}$ \\
		\end{tabular}
	\end{ruledtabular}
\end{table}

\section{\label{sec:spectra}Speed-dependent spectra of the optimal driving pulse}

Due to the time-frequency uncertainty relation, the minimal frequency bandwidth of a pulse should be inversely proportional to the pulse duration. Thus, the spectral density should also be different when varying the gate time. 
Although this is true in terms of the overall shape, we show that the Fourier component at the qubit frequency is nearly constant of gate speed.

\begin{figure}
	\centering
	\includegraphics[width=1\linewidth]{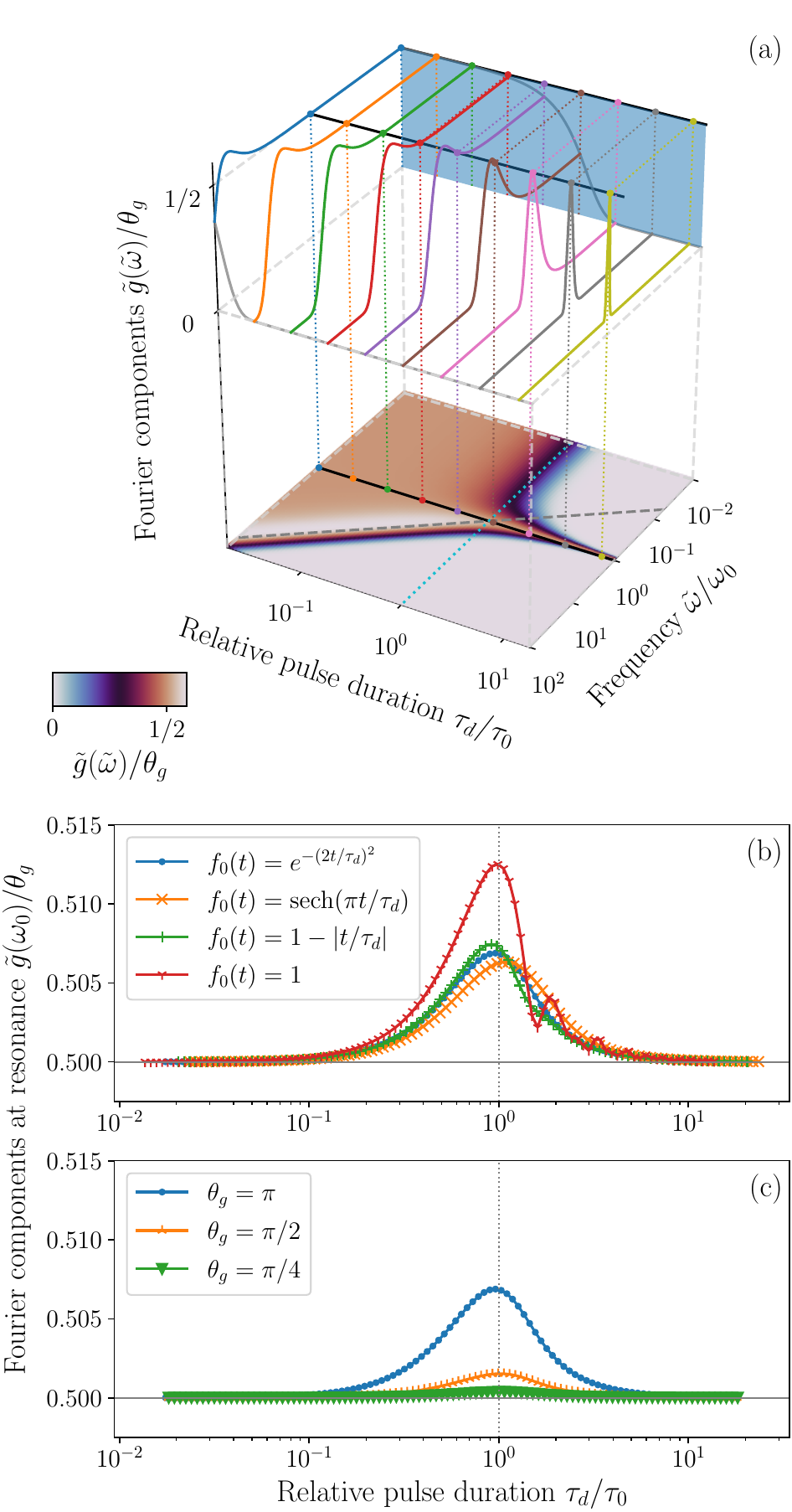}
	\caption{Speed-dependent spectra of optimal pulses. (a) Fourier components $\tilde{g}(\tilde{\omega})$ normalized by the gate rotation angle $\theta_{g}$ in dependence on the pulse duration. The envelope function is Gaussian $f_{0}(t) = e^{-(2t/\tau_{d})^{2}}$. The dotted line at the bottom indicates the transition pulse duration $\tau_{d} = \tau_{0}$ [see Eq.\,(\ref{eq:tau0}) and Table \ref{table:tau0}]. The black solid lines at $\tilde{\omega} = \omega_{0}$ show the Fourier components at resonance, i.e., $\tilde{g}(\omega_{0})$, which is close to $\theta_{g}/2$. (b) Fourier components at resonance for four different envelope functions $f_{0}(t)$, namely, Gaussian, hyperbolic secant, triangular, and constant function. The rotation angle is fixed as $\theta_{g} = \pi$. (c) Fourier components at resonance for gate angles $\theta_{g}=\pi,\pi/2$, and $\pi/4$. The envelope function is Gaussian.}
	\label{fig:spectra-and-envelope-comparison}
\end{figure}

In Fig.\,\ref{fig:spectra-and-envelope-comparison}(a), we plot the Fourier components $\tilde{g}(\tilde{\omega}) = \int_{-\infty}^{\infty}{dt \Omega f(t)e^{-i\tilde{\omega} t}}$ of the optimal pulse shapes for several pulse durations. The envelope function is set to be Gaussian as in Figs.\,\ref{fig:pulse-shapes-and-bloch-vectors} and \ref{fig:average-gate-fidel-and-parameters}. Since we use an even pulse, the Fourier components are real. Also, the pulse envelope function $f_{0}(t)$ is zero outside of the interval $[-T/2,T/2]$. Thus, we can write $\tilde{g}(\tilde{\omega}) = \int_{-T/2}^{T/2}{dt \Omega f(t)\cos(\tilde{\omega} t)}$.
We note that in the multicycle regime ($\tau_{d} > \tau_{0}$) the maximum Fourier component appears nearly at the resonance, $\tilde{\omega}_{\mathrm{max}} \simeq \omega_{0}$, 
whereas in the subcycle regime ($\tau_{d} < \tau_{0}$)
the peak-value frequency gets blue-detuned approximately by a factor of $\tau_{0}/\tau_{d}$, i.e., $\tilde{\omega}_{\mathrm{max}} \simeq (\tau_{0}/\tau_{d}) \omega_{0}$.
This is consistent with the speed-dependent optimal central frequency $\omega$ shown in Fig.\,\ref{fig:average-gate-fidel-and-parameters}(c).

The Fourier component at resonance, $\tilde{g}(\omega_{0})$, however, is nearly constant throughout the whole gate speed range.
In Fig.\,\ref{fig:spectra-and-envelope-comparison}(a), the upper black solid line at the resonant frequency, $\tilde{\omega} = \omega_{0}$, indicate the Fourier components at $\omega_{0}$, which are normalized by the gate rotation angle $\tilde{g}(\omega_{0})/\theta_{g}$. 
We see that 
$\tilde{g}(\omega_{0}) \simeq \theta_{g}/2$ 
in both multicycle and subcycle regimes. 
This can be shown analytically in each limit. In the subcycle limit, where the gate time $T$ is sufficiently short compared to the qubit period $T_{0} = 2\pi/\omega_{0}$, we have $\cos(\omega_{0}t) = 1 + \mathcal{O}[(\omega_{0}T)^{2}] \simeq 1$. The Fourier component becomes 
\begin{equation}\label{eq:fourier-resonant-subcycle}
	\tilde{g}(\omega_{0}) \xrightarrow{\tau_{d} \ll T_{0}} \int_{-T/2}^{T/2} dt \Omega f(t) = \frac{\theta_{g}}{2},
\end{equation}
where the last equality holds due to Eq.\,(\ref{eq:theta-theta_x}). See also the pulse on the left in Fig.\,\ref{fig:average-gate-fidel-and-parameters}(b). In the multicycle limit, where $\tau_{d} \gg T_{0}$ and the driving frequency is resonant to the qubit, $\omega = \omega_{0}$, we have $f(t) = f_{0}(t) \cos(\omega_{0}t)$ and thus $\tilde{g}(\omega_{0}) = \int_{-T/2}^{T/2}{dt \Omega f_{0}(t) \cos^{2}(\omega_{0}t)}$. Since $\cos^{2}(\omega_{0}t) = (1/2)[1 + \cos(2\omega_{0}t)]$ and $\Omega \ll \omega_{0}$, the integral is approximated, for a slowly-varying envelope function $f_{0}(t)$, as
\begin{equation}\label{eq:fourier-resonant-multicycle}
	\tilde{g}(\omega_{0}) \xrightarrow{\tau_{d} \gg T_{0}} \int_{-T/2}^{T/2}{dt \Omega f_{0}(t)\frac{1}{2}} = \frac{\theta_{g}}{2},
\end{equation}
where the last equality is ensured by Eq.\,(\ref{eq:Omega-tau-d-rwa}), 
cf. also the pulse on the right in Fig.\,\ref{fig:average-gate-fidel-and-parameters}(b). The two limits, Eqs.\,(\ref{eq:fourier-resonant-subcycle}) and (\ref{eq:fourier-resonant-multicycle}), are valid for general nonnegative monotonic (in $|t|$) envelope functions $f_{0}(t)$ and angles $\theta_{g} \in [0,\pi]$. In Figs.\,\ref{fig:spectra-and-envelope-comparison}(b) and \ref{fig:spectra-and-envelope-comparison}(c), we compare Fourier components at resonance for different envelope functions and gate rotation angles, respectively. 
We see that the ratio between the Fourier component at the qubit frequency and the gate angle, i.e., $\tilde{g}(\omega_{0})/\theta_{g}$, converges in both subcycle and multicycle limits to a constant, namely, $1/2$, for different envelope functions $f_{0}(t)$ and gate angles $\theta_{g}$. 
The deviation from $1/2$ in the intermediate regime ($\tau_{d} \sim \tau_{0}$) is $1\sim3\%$ for the envelopes and gate angles shown in Fig.\,\ref{fig:spectra-and-envelope-comparison}. In Sec.\,\ref{sec:rot-frame-magnus}, we explain this deviation.

\section{\label{sec:rot-frame-magnus}Rotating-frame Magnus expansion}

So far, we have presented expansions for the rotating-frame propagator in the subcycle and the multicycle limits. In the former, the expansion is performed with respect to $\tau_{d}/T_{0}$, since although the driving strength is strong ($\Omega \gg \omega_{0}$), the pulse duration is short ($\tau_{d} \ll T_{0}$).
In the latter, the small parameter is $\Omega / \omega_{0}$, due to the long ($\tau_{d} \gg T_{0}$) but weak ($\Omega \ll \omega_{0}$) driving.
Each expansion works well in the corresponding regime, but it may break down for a gate time which cannot be attributed to any of these regimes. The corresponding intermediate regime is hard to access with sufficient accuracy of either expansion.
In both considered regimes, however, the product of the driving strength and the pulse duration is bounded since
\begin{equation*}
	(\Omega/\omega_{0})\times(\tau_{d}/T_{0}) = 
	\frac{\Omega\tau_{d}}{2\pi},
\end{equation*}
for a given gate angle $\theta_{g}$.
The effective pulse area $\Omega\tau_{d}$ converges to a constant in the subcycle limit, Eq.\,(\ref{eq:Omega-tau-d-subcycle}), and to another constant in the multicycle limit, Eq.\,(\ref{eq:Omega-tau-d-rwa}).
Therefore, we aim to expand the exponent of the propagator $U_{\mathrm{rot}}$ in terms of $\Omega \tau_{d}$ and find optimal values of the parameters analytically for all gate time regimes.

We first note that the even pulse shape $f(-t)=f(t)$, by setting the carrier-envelope phase $\phi=0$, implies the time-reversal symmetry for the Hamiltonian, $H_{\mathrm{rot}}(-t) = H_{\mathrm{rot}}^{*}(t)$. The complex conjugate on $H_{\mathrm{rot}}$ is done by conjugating the matrix elements in the standard basis, for which $\sigma_{x}^{*}=\sigma_{x}$ and $\sigma_{y}^{*}=-\sigma_{y}$. One consequence of this symmetry is that the propagator at the final time is identical to its transpose, i.e., 
\begin{equation*}
	U_{\mathrm{rot}}(T/2,-T/2) \equiv U_{\mathrm{rot}} = U_{\mathrm{rot}}^{\mathrm{T}}.
\end{equation*}
For a two-level system, this implies the propagator has only $\sigma_{x}$ and $\sigma_{z}$ components in the exponent, since $\sigma_{y}^{\mathrm{T}} = -\sigma_{y}$. This has already been used in Sec.\,\ref{sec:model-single-qubit-gates} to satisfy Eq.\,(\ref{eq:s1c-zero}). Using this symmetry, we directly write the propagator as
\begin{equation*}
	U_{\mathrm{rot}} \equiv \exp[-i(A^{x}\sigma_{x}/2 + A^{z}\sigma_{z}/2)],
\end{equation*}
where the coefficients $A^{x}$ and $A^{z}$ depend on the driving pulse parameters.
Thus, setting 
\begin{subequations}
	\begin{align}
		\label{eq:Ax-theta_g}
		A^{x} &= \theta_{g},\\
		\label{eq:Az-zero}
		A^{z} &= 0
	\end{align}
\end{subequations}
gives our target gate $R_{x}(\theta_{g}) = \exp[-i\theta_{g}\sigma_{x}/2]$.

In order to find the optimal values of the parameters, we obtain analytical expressions for $A^{x}$ and $A^{z}$ by using the Magnus expansion up to the third order with respect to the effective pulse area $\Omega\tau_{d}$,
\begin{equation*}
	A^{j} \equiv \sum_{n=1}^{3}{A^{j}_{n}} + \mathcal{O}[(\Omega\tau_{d})^{4}],
\end{equation*}
where $j \in \{x,z\}$.
The expressions have also been presented in Ref.\,\cite{magnus-few-cycle} for a few-cycle pulse with a general carrier-envelope phase $\phi$. Here, we are interested in the case of even pulse ($\phi = 0$), which simplifies the expressions. The simpler expressions are useful since we want to inverse-engineer the fidelity to find optimal parameters for implementing a quantum gate. To write the expressions, we first define auxiliary functions as
\begin{subequations}
\begin{align*}
	\theta_{1}(t,0) &= \int_{0}^{t}{dt' \Gamma_{x}(t')},\\
	\theta_{n}(t,0) &= \int_{0}^{t}{dt' \Gamma_{y}(t') \theta_{n-1}(t',0)},
\end{align*}
\end{subequations}
for $n \ge 2$. Here, $\Gamma_{x}$ and $\Gamma_{y}$ are components of the Hamiltonian, Eq.\,(\ref{eq:H-rot}),
\begin{equation*}
	H_{\mathrm{rot}}(t) \equiv \hbar \mathbf{\Gamma}(t)\cdot\pmb{\sigma}/2,
\end{equation*}
i.e., $\Gamma_{x}(t) + i\Gamma_{y}(t) \equiv 2\Omega f(t) e^{-i\omega_{0} t}$. Then, the Magnus terms are given as
\begin{equation}\label{Ax-upto-3rd}
	A^{x} = A_{1}^{x} + A_{3}^{x} + \mathcal{O}[(\Omega\tau_{d})^{5}]
\end{equation}
with $A_{1}^{x} = 2\theta_{1}(T/2,0)$ and $A_{3}^{x} = -2\theta_{3}(T/2,0)$,
and
\begin{equation}\label{eq:Az-upto-3rd}
	A^{z} = A_{2}^{z} + \mathcal{O}[(\Omega\tau_{d})^{4}]
\end{equation}
with $A_{2}^{z} = -2\theta_{2}(T/2,0)$.
If we consider terms up to the second order in $\Omega\tau_{d}$, we have 
\begin{subequations}
\begin{align}
	\label{eq:Ax-cond-2nd-order}
	A_{1}^{x} &= \theta_{g},\\
	A_{2}^{z} &= 0.
\end{align}
\end{subequations}
Considering terms up to the third order in $\Omega\tau_{d}$, we have 
\begin{subequations}
	\begin{align}
		\label{eq:Ax-cond-3rd-order}
		A_{1}^{x} + A_{3}^{x} &= \theta_{g},\\
		A_{2}^{z} &= 0.
	\end{align}
\end{subequations}
In both cases, we have $A_{2}^{z} = 0$. Since $A_{2}^{z}$ is proportional to $(\Omega\tau_{d})^{2}$ and $A_{2}^{z}/(\Omega\tau_{d})^{2}$ is independent of $\Omega$, $A_{2}^{z} = 0$ determines $\omega\tau_{d}$, from which the central frequency is identified.
Then, we insert $\omega$ into Eq.\,(\ref{eq:Ax-cond-2nd-order}) or Eq.\,(\ref{eq:Ax-cond-3rd-order}) to obtain $\Omega\tau_{d}$ within the second- or third-order Magnus expansion.

\begin{figure}
	\centering
	\includegraphics[width=1\linewidth]{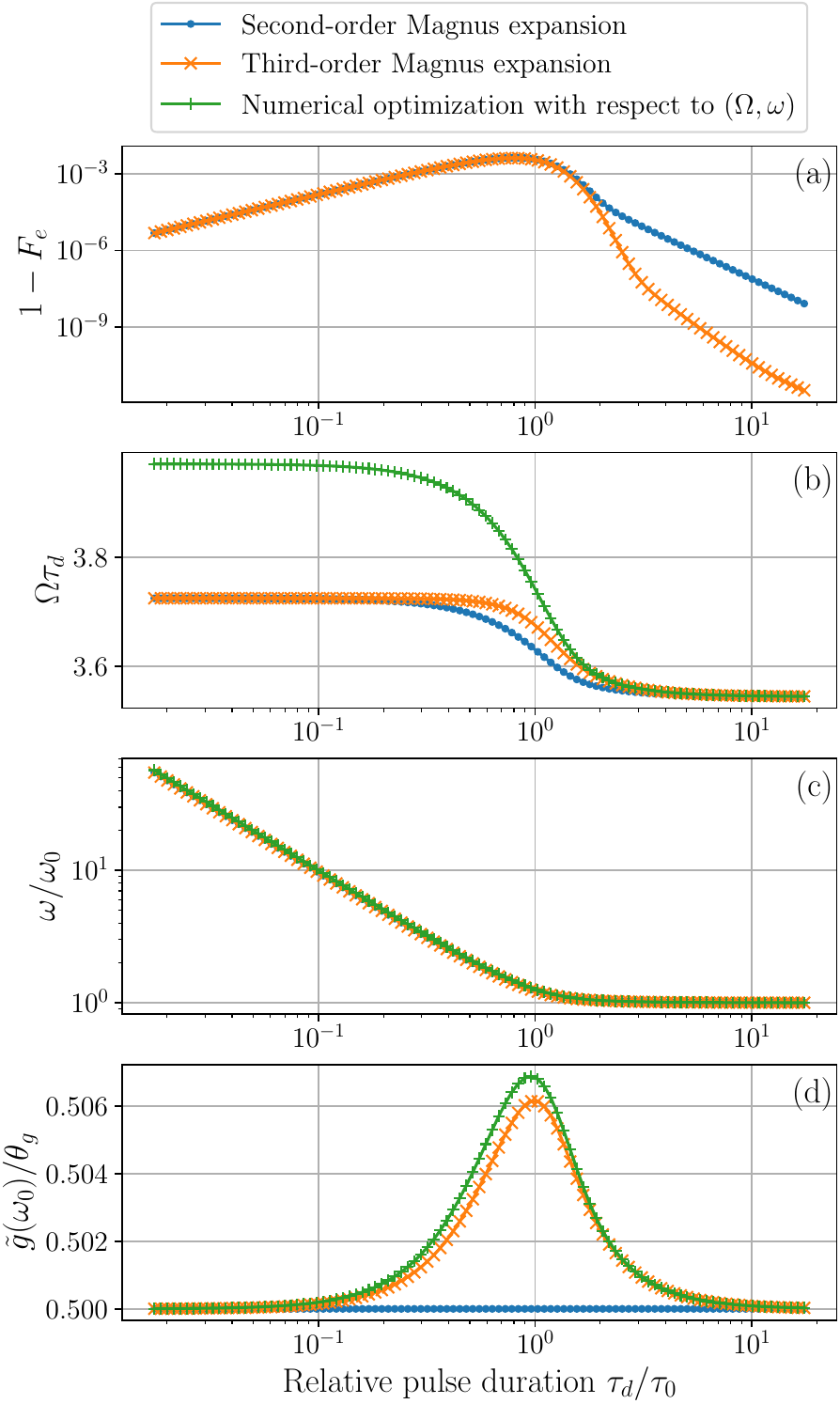}
	\caption{Comparison between optimal $\Omega$ and $\omega$ obtained by the Magnus expansion and by the numerical optimization of fidelity. The pulse envelope function is Gaussian and the gate angle is $\theta_{g}=\pi$. (a) Entanglement fidelity $F_{e}$. The fidelity from the numerical optimization is not visible in this log-log plot since $F_{e}=1$ up to numerical precision. (b,c) Dependence of the normalized optimal values $\Omega\tau_{d}$ and $\omega/\omega_{0}$ on the pulse duration. The numerically obtained values are from Fig.\,\ref{fig:average-gate-fidel-and-parameters}(b) and \ref{fig:average-gate-fidel-and-parameters}(c). (d) Fourier component at the qubit frequency, $\tilde{g}(\omega_{0})$, normalized by the gate angle $\theta_{g}$. The data from the numerical optimization are the same as in Fig.\,\ref{fig:spectra-and-envelope-comparison} for $f_{0}(t)=e^{-(2t/\tau_{d})^{2}}$ and $\theta_{g}=\pi$.}
	\label{fig:magnus-expansion-params-fidel-fourier}
\end{figure}

In Figs.\,\ref{fig:magnus-expansion-params-fidel-fourier}(b) and \ref{fig:magnus-expansion-params-fidel-fourier}(c), we compare the values of the parameters obtained from the second-order and the third-order Magnus expansion with those obtained by the numerical optimization shown in Figs.\,\ref{fig:average-gate-fidel-and-parameters}(b) and \ref{fig:average-gate-fidel-and-parameters}(c). We see that in the multicycle limit, the values obtained by the Magnus expansion converge to the results from the numerical optimization. In the subcycle limit, however, there are some difference between the two.
The corresponding infidelity $1-F_{e}$ is shown in Fig.\,\ref{fig:magnus-expansion-params-fidel-fourier}(a), where $F_{e}$ is the entanglement fidelity introduced in Sec.\,\ref{sec:model-single-qubit-gates}.

We further note that the Fourier component at the qubit frequency can be written as
\begin{equation*}
	\tilde{g}(\omega_{0}) = A_{1}^{x}/2.
\end{equation*}
Thus, from Eqs.\,(\ref{eq:Ax-cond-2nd-order}) and (\ref{eq:Ax-cond-3rd-order}), we have
\begin{subequations}
\begin{align}
	\label{eq:fourier-magnus-2nd}
	\tilde{g}(\omega_{0}) &= \theta_{g}/2\,\text{(second order)},\\
	\label{eq:fourier-magnus-3rd}
	\tilde{g}(\omega_{0}) &= \theta_{g}/2 - A_{3}^{x}/2\,\text{(third order)},
\end{align}
\end{subequations}
respectively. We see that the deviation of the Fourier component at the qubit frequency from $\theta_{g}/2$, which was observed in Figs.\,\ref{fig:spectra-and-envelope-comparison}(b) and \ref{fig:spectra-and-envelope-comparison}(c), can be predominately attributed to $-A_{3}^{x}/2$. In Fig.\,\ref{fig:magnus-expansion-params-fidel-fourier}(d), we compare the resonant Fourier component normalized by the gate angle, $\tilde{g}(\omega_{0})/\theta_{g}$, obtained from the numerical optimization to that from the second- and third-order Magnus expansion, given by Eqs.\,(\ref{eq:fourier-magnus-2nd}) and (\ref{eq:fourier-magnus-3rd}), respectively.

\section{\label{sec:universal-set}Construction of a universal set of single-qubit gates}
We now comment on constructing a universal set of single-qubit gates using the pulse parameters for $R_{x}(\theta_{g})$.
It is sufficient to find a way to implement $R_{y}(\theta_{g})$ for $-\pi \le \theta_{g} \le \pi$. 
The parameters are the same as those for $R_{x}(\theta_{g})$, except that the pulse is translated in time by an odd multiple of $T_{0}/4$. 
This is due to the fact that this operates on a rotating frame.
To see this, let us consider a driving strength $\Omega_{\theta}$ and the pulse shape $f_{\theta}(t)$ centered at $t=0$ which generate $R_{x}(\theta)$. Translate the pulse in time by $t_{0}$ to have $\Omega f(t) = \Omega_{\theta} f_{\theta}(t-t_{0})$ so that the pulse drives the qubit for $t\in[t_{0}-T/2,t_{0}+T/2]$. Set $t' \equiv t - t_{0}$ and note that from Eq.\,(\ref{eq:H-rot}) follows
\begin{equation*}
	\begin{split}
		H_{\mathrm{rot}}(t) 
		&= H_{\mathrm{rot}}(t_{0} + t')\\
		&= R_{z}^{\dagger}[\omega_{0}(t_{0}+t')] \{\hbar \Omega_{\theta} f_{\theta}(t') \sigma_{x}\} R_{z}[\omega_{0}(t_{0}+t')]\\
		&= R_{z}^{\dagger}(\omega_{0}t_{0}) H_{\theta}(t') R_{z}(\omega_{0}t_{0}).
	\end{split}
\end{equation*}
where $H_{\theta}(t) \equiv \hbar \Omega_{\theta} f_{\theta}(t) [\cos(\omega_{0}t) \sigma_{x} - \sin(\omega_{0}t) \sigma_{y}]$ generates $R_{x}(\theta)$.
Inserting this expression into the Schr\"{o}dinger equation gives
\begin{equation*}
	U_{\mathrm{rot}}(t_{0}+T/2,t_{0}-T/2;0) = R_{z}^{\dagger}(\omega_{0}t_{0}) R_{x}(\theta) R_{z}(\omega_{0}t_{0}).
\end{equation*}
Thus, for an integer $k$, a delay of $t_{0} = (2k)T_{0}/4$ and $(2k+1)T_{0}/4$ gives $R_{x}[(-1)^{k}\theta]$ and $R_{y}[(-1)^{k+1}\theta]$, respectively. For example, $R_{y}(\pi/2)$ can be generated by $t_{0} = T_{0}/4$ with $\Omega = -\Omega_{\pi/2}$ or $t_{0} = 3T_{0}/4$ with $\Omega = \Omega_{\pi/2}$. We note that a delay of $T_{0}/4$ corresponds to a phase offset of $\omega_{0}T_{0}/4$, which equals to $\pi / 2$. Under the RWA, where the control field $f(t)$ contains multiple cycles, it is possible to induce such a shift through the phase of $f(t)$ by $\phi \rightarrow \phi + \pi/2$ in Eq.\,(\ref{eq:pulse-shape-ft}) since the gate fidelity is sufficiently insensitive to the carrier envelope phase $\phi$ \cite{PhysRevA.96.022330}. Even a few-cycle pulse for a gate time $T \simeq 4 T_{0}$ exhibits low sensitivity to the phase \cite{PRXQuantum.5.020326}. For a faster gate where $T$ is comparable or shorter than the qubit period $T_{0}$, the carrier-envelope phase $\phi$ should be fixed and a precise time delay is needed. 

%

We note that rotations around two different axes are sufficient for generating a universal set of single-qubit gates. For example, $R_{x}(\theta_{g})$ and $R_{y}(\theta_{g})$ can be selected where $\theta_{g} \in [-\pi,\pi]$.
The $R_{z}$ gates can be synthesized by the $R_{x}$ and $R_{y}$ gates, but they can also be realized virtually, which does not take time but has an equivalent effect by adjusting the subsequent gate sequence \cite{PhysRevA.96.022330}.

\section{\label{sec:conclusion}Conclusion}

We have shown that two parameters, the Rabi frequency $\Omega$ and the central frequency $\omega$, are sufficient to parameterize optimal pulse shapes for building a universal set of single-qubit gates at an arbitrary timescale with unit fidelity.
In order to obtain the optimal values of the parameters analytically, we used unitary perturbation theories with respect to different small quantities in which we expand.
In the long gate time limit (or, the multicycle limit defined by $T \gg T_{0}$), 
we used $\Omega/\omega_{0} \ll 1$ for the expansion. The RWA corresponds to the zeroth-order terms \cite{ZEUCH2020168327}. The perturbation expansion broke down when the driving is strong, $\Omega \gtrsim \omega_{0}$, which is inevitable for a fast gate such that $T \lesssim T_{0}$, due to the quantum speed limit, Eq.\,(\ref{eq:speed-limit}). 
However, in the short gate time limit (or, the subcycle limit defined by $T \ll T_{0}$), we used $T/T_{0} \ll 1$ instead as the quantity in which we expand. 
By using $\Omega/\omega_{0}$ and $T/T_{0}$ as the small quantities for the expansions in the long- and short-$T$ limits, we obtained the optimal values of the pulse parameters with an error $\mathcal{O}[(\Omega/\omega_{0})^{1}]$ and $\mathcal{O}[(T/T_{0})^{2}]$, respectively.
To further reduce the error, we performed a numerical optimization of the gate fidelity with respect to $\Omega$ and $\omega$. 
We started from a gate time deep in the subcycle regime and found optimal parameters achieving unit fidelity up to numerical precision.
Instead of having a random guess, we used the set of analytically obtained parameters as an initial condition of the numerical optimization. 
Then, we inductively optimized the fidelity by gradually increasing the gate time to find optimal parameters for all $T$ regimes, including the intermediate ($T \sim T_{0}$) and the multicycle regimes.

In an experiment, there can be a preferred pulse shape. 
For example, a pulsed laser can naturally have hyperbolic secant ($\mathrm{sech}$) shape \cite{PhysRevLett.18.908,*PhysRev.183.457}. If the driving uses a Raman process, the pulse shape becomes its square ($\mathrm{sech}^{2}$) \cite{PhysRevLett.105.090502}. For a system with a limited anharmonicity or a dense spectrum, the Gaussian pulse shape is preferred to suppress residual excitation \cite{PhysRevLett.103.110501,PhysRevLett.116.020501,PhysRevApplied.22.014007}.
We report that optimal parameters with unit fidelity exist at arbitrary gate time for different pulse envelope functions $f_{0}(t)$, including the Gaussian, hyperbolic secant, triangular, and the constant function.
We also note that the carrier part of the pulse can be replaced by other types of oscillating functions. For example, a triangular wave (i.e., zigzag) can be useful to generalize the finite-time Landau-Zener driving \cite{PhysRevA.53.4288} and the corresponding spin-echo-type driving \cite{PhysRevX.11.011010} to implement an arbitrary-speed quantum gate.

We found that the Fourier component of the optimal pulse shape at the qubit frequency converges to the same value, namely, the half of the gate rotation angle, in both limits.
We observed that the resonant Fourier component increases by few percents in the intermediate gate time regime ($\tau_{d} \sim \tau_{0}$).
We obtained an approximate analytical expression for that small increase by expanding the generated unitary operator with respect to the effective pulse area $\Omega\tau_{d} \sim \theta_{g}$. The dominant part of the increase near the transition pulse duration $\tau_{0}$ has been attributed to the third-order Magnus term $A_{3}^{x} = \mathcal{O}[(\Omega\tau_{d})^{3}]$, which vanishes both in the subcycle and in the multicycle limit.

We conclude with remarks on potential experimental noises expected in each gate time regime.
In the subcycle regime, since the gate time is short and the pulse is strong, the intensity noise can represent the dominating effect on the gate fidelity \cite{Chew2022-mz}, whereas in the multicycle regime, the phase noise \cite{PhysRevLett.121.123603} can be the limiting factor. The errors due to these noises can be mitigated by improving the driving pulse sources \cite{mahesh2024generation480nmpicosecond,denecker2024measurementfeedforwardcorrectionfast} or by using a special pulse shape which is robust to the intensity and/or the phase error \cite{PhysRevLett.111.050404}.
\begin{acknowledgments}
S.A., K.P., and D.C. were supported by the KAIST Venture Research Program for Graduate \& Ph.D students.
S.A. and A.S.M. were supported by the NRF grant funded by the Korean government (MSIT) under Grant No. 2020R1A2C1008500. T.C. acknowledges support from the NRF under Grant No. RS-2022-NR068814, RS-2024-00442855, and RS-2024-00466865.
\end{acknowledgments}

\bibliography{manuscript}

\end{document}